\documentclass[12pt,a4paper]{article}

\usepackage[english]{babel}
\usepackage[T2A]{fontenc}
\usepackage[cp1251]{inputenc}
\usepackage{amsmath}
\usepackage{graphicx}
\usepackage{amssymb}
\usepackage{color}
\usepackage{amsfonts}
\usepackage{wrapfig}
\usepackage{caption}

\usepackage{hyperref}
\usepackage{physics}

\begin{document}
	
	\binoppenalty=10000
	\relpenalty=10000

\begin{center}
	\textbf{\Large{Influence of microscopic parameters on phase behavior of a cell model with Curie-Weiss interaction}}
\end{center}

\begin{center}
	O.A.~Dobush$^*$ \footnote{e-mail:  dobush@icmp.lviv.ua}, M.P.~Kozlovskii$^*$, I.V.~Pylyuk$^*$, Yu.O. Plevachuk$^{\dag,\ddag}$
\end{center}

\begin{flushleft}
	$^*$ Institute for Condensed Matter Physics of the National Academy of Sciences of
	Ukraine \\ 1 Svientsitskii Street, 79011 Lviv, Ukraine \\	
	$^\dag$ Ivan Franko National University of Lviv, Department for Metal Physics, 8 Kyrylo and Mephodiy Street, 79005 Lviv, Ukraine \\	
	$^\ddag$ Institute of Physics, Slovak Academy of Sciences, 9 D\'{u}bravsk\'{a} cesta, 84511 Bratislava, Slovakia
\end{flushleft}

\vspace{0.2cm}

\begin{center}
	\small \textbf{Abstract} 
\end{center}

\small 
We investigate how varying two microscopic parameters \--- cell volume and the ratio between repulsion and attraction intensities \--- affect the phase behavior of a cell model with Curie-Weiss-type interaction. The analysis is based on an exact solution previously derived for this model in the grand canonical ensemble. At sufficiently low temperatures, the cell model exhibits multiple first-order phase transitions. By varying the cell volume and the repulsion-to-attraction ratio, we represent a quantitative comparison of the chemical potential and pressure isotherms, along with the pressure-temperature and temperature-density phase diagrams. Our results demonstrate that altering these microscopic parameters induces quantitative changes in the phase diagrams of the cell model. 

\vspace{0.2cm}

	\normalsize 
\section{\label{sec:model} Introduction}

The analysis of fluid models using simplified interaction potentials is essential for understanding the connection between microscopic interactions and macroscopic phase behavior.  Fluid phase diagrams can be significantly altered by even small changes in interaction parameters, such as well depth, range, or softness of repulsion as shown by computer modeling and theoretical investigations employing statistical mechanics, integral equations, and perturbation theories \cite{hm113}. Computer simulation studies demonstrate that decreasing the range of attraction in a system of hard spheres with an attractive Yukawa potential~\cite{hf194,d102} as well as in Lennard-Jones like models~\cite{wrdf120,ahp124} leads to the vanishing of stable fluid-fluid transitions and remaining stable fluid-solid transitions. Moreover, reducing the interaction range flatten out the vapor-liquid coexistence curves and shrinks the range of temperatures for a stable liquid phase in attractive hard-core Yukawa fluids, as shown by computer modeling~\cite{gta101,glo108,s109}, reference hypernetted chain integral equation theory~\cite{la194}, and perturbative density functional theory~\cite{sbt121}. The same effect is also observed in square-well fluids when decreasing the range of the potential well~\cite{vmrjm192}, in systems with interactions modeled by the approximate non-conformal theory potentials when weakening the attractive softness~\cite{fmgli116}, in Mie fluids when either increasing repulsion exponent~\cite{wshh117,su123} or increasing dispersive exponent~\cite{su123}, in fluids with modified Buckingham exponential-6 potential when increasing repulsion intensity~\cite{ep198,p100}.

The aim of this study is to demonstrate how changes in the microscopic parameters of the continuous Curie-Weiss model affect its phase behavior.

This work expands on the earlier research \cite{kd122}, which is based on a cell model with Curie-Weiss interaction. In this approach, an open system of interacting particles is considered in the volume $V$ divided into $N_v$ cells. Each cell is of volume $v=V/N_v = c^3$, where $c$ represents the linear size of a cell.

It is important to note that, unlike the cell gas model — where each cell can contain at most one particle or remain empty \cite{rebenko_13,rebenko_15} — our model allows multiple particles to occupy a single cell. The cell model introduced in \cite{kkd118,kkd120} has been previously applied to the microscopic description of the critical behavior of Morse fluids \cite{kpd118,p120,pkdd123}.

A key advantage of this model is that it has an exact solution in the case of Curie-Weiss interaction \cite{kkd118, kkd120}. One of its notable features is its complex phase behavior: it possesses multiple critical points and exhibits a sequence of first-order phase transitions at sufficiently low temperatures. The mathematical formulation is carried out within the framework of the grand canonical ensemble (see, for example, \cite{hm113}).

The grand partition function of the continuous system is as follows \cite{hm113,hmo156}
\begin{equation}
	\Xi = \sum_{N=0}^\infty \frac{\zeta^N}{N!} \int \limits_V \mathrm dx_1 ...
	\int \limits_V \mathrm dx_N \exp\left[ -\frac{\beta}{2} \sum_{x,y\in\gamma} \Psi_{N_v} (x,y)\right],
	\label{1d1cw}
\end{equation}
where $\zeta =\exp(\beta\mu)/\Lambda^3$ is the activity, $\mu$ is the chemical potential, $\beta = (k_BT)^{-1}$ is the inverse temperature, $\Lambda$ is the thermal de Broglie wavelength, $\gamma=\{x_1,...,x_N\}$, $x_i$ are the coordinates of the $i$-th particle. Like \cite{kd122,kkd118,kkd120}, we set $\Lambda=1 $. Henceforth, we choose the Curie-Weiss potential as the interaction potential $\Psi_{N_v} (x,y)$ \cite{kd122,kkd120}:
\begin{equation}\label{1d2cw}
	\Psi_{N_v} (x, y) = - \frac{g_a}{N_v} + g_r \sum_{l =1}^{N_v}
	I_{\Delta_l}(x) I_{\Delta_l}(y).
\end{equation}
Here
\begin{equation}
	I_{\Delta_l}(x) = \left\{
	\begin{array}{ll}
		1, & x\in \Delta_l \\
		0, & x\notin \Delta_l
	\end{array}
	\right .
	\label{1d3cw}
\end{equation}
is the characteristic function, which determines if a particle at $x$ belongs to a cubic cell, $\Delta_l = (-c/2,c/2]^3 \subset \mathbb{R}^3$. The term with $g_a>0$ describes attraction, which is taken equal for all particles. The term with $g_r>0$ describes repulsion between two particles contained in one and the same cell.
The stability of the interaction \cite{r170} is ensured by the inequality $g_r>g_a>0$.

In the following Section~\ref{sec:eos}, we recall the basic relations and present the explicit expression for the equation of state of the cell model with the Curie-Weiss interaction. In two subsequent sections, we study the effect of changing two microscopic parameters in the model on its phase behavior: the cell volume (Section~\ref{sec:volume}) and ratio between repulsion and attraction intensities (Section~\ref{sec:interaction}). Final discussion on the role of these key parameters is highlighted in Conclusions. 

\section{\label{sec:eos} Equation of state and other known relations}

In~\cite{kd122}, the grand partition function (\ref{1d1cw}), in the case of interaction (\ref{1d2cw}), is defined as
\begin{equation}
\Xi = (N_v/(2\pi p_a))^{1/2} \int \mathrm d z \exp(N_v E_0(z, \mu)),
\label{1d4cw}
\end{equation}
where
\begin{equation}
E_0(z, \mu) = - \frac{1}{ 2 p_a}(z - \beta\mu-\ln v^*)^2 + \ln K_0(z).
\label{1d5cw}
\end{equation}
In this work, the dimensionless volume $v^* = v/\Lambda^3$ is taken equal to $v$, since the de Broglie wavelength, $\Lambda$, was assumed to be equal to 1. The special functions
\begin{equation}
K_m(z) = \sum_{n=0}^{\infty} \frac{1}{n!} n^m
\exp\left( z n -\frac{fp_a}{2}n^2   \right)
\label{1d6cw}
\end{equation}
depend on the repulsion intensity, $g_r$, and the temperature, since
\begin{align}
&
f = g_r/g_a, \quad p_a=\beta g_a, \quad fp_a = p_r, \quad
p_r=\beta g_r, \nonumber \\
&
\beta = \beta_c/(1 + \tau), \quad \tau = (T - T_c)/T_c.
\label{1d7cw}
\end{align}

The expression (\ref{1d4cw}) is calculated in \cite{kd122} by the Laplace method \cite{f189}. An asymptotic solution at the thermodynamic limit ($N_v \to \infty$, $V \to \infty$, $V/N_v=v=const$) gives the grand partition function in the following form:
\begin{equation}
	\Xi = \exp \left ( N_v E_0(z, \mu)\right )\big|_{z=\bar z}.
	\label{1d8cw}
\end{equation}
The quantity $\bar z$ can be determined from the condition of extrema $\partial E_0(z,\mu)/ \partial z = 0$. The explicit expression for the chemical potential, $\mu$, as a function of $\bar z$ has the form
\begin{equation}\label{1d10cw}
\beta\mu = \bar z - p_a \frac{K_1(\bar z)}{K_0(\bar z)}  - \ln v^*.
\end{equation}

Using the relation for the average number of particles $\left< N \right> = \partial\ln\Xi/\partial\beta\mu$, we arrive at the following expression for the reduced particle number density, $\bar n = \left< N \right> / N_v$,
\begin{equation}\label{1d11cw}
\bar n = \frac{K_1(\bar z)}{K_0(\bar z)},
\end{equation}
which literally means the average number of particles per one cell.

Combining the equations \eqref{1d5cw} and \eqref{1d10cw}, we exclude the chemical potential from the function $E_0(z, \mu)$ and rewrite it in the following form:
\begin{equation}
E_0(\bar z) = \ln K_0(\bar z) - \frac{p_a}{2}\left(\frac{K_1(\bar z)}{K_0(\bar z)}\right)^2.
\end{equation}
Then, the equation \eqref{1d8cw} can be represented as $\Xi = \exp \left ( N_v E_0(\bar z)\right )$. 

The continuous Curie-Weiss model exhibits a sequence of phase transitions with different critical temperatures \cite{kd122}. The solutions of the system of equations
\begin{equation}
\left\{
\begin{array}{ll}
	& \cfrac{d\mu}{d z} \Big|_{z=\bar z} = 0, \\
	& \cfrac{d^2\mu}{dz^2} \Big|_{z=\bar z} = 0
\end{array}
\right .
\label{1d12cw}
\end{equation}
give the critical values of the parameter $p_a$ and the quantity $\bar z$ for $n$ phase transitions. Corresponding formula for calculating the critical temperature, $T_c^{(n)}$, is
\begin{equation}
k_B T_c^{(n)} = g_a / p_{ac}^{(n)}.
\label{1d13cw}
\end{equation}      
The specific values of the critical temperatures depend on the choice of the attraction intensity,~$g_a$. Currently, we are more interested in the overall picture. Therefore, there is no need in assigning any particular value to $ g_a $. In addition, to simplify the notations, we assume that
\begin{equation}
T_c^{(1)} = T_c, \qquad \beta_c^{(1)} = \beta_c.
\label{1d14cw}
\end{equation}
According to \cite{kd122,kkd120}, solving the system of equations
\begin{equation}
\left\{
\begin{array}{ll}
	& E_0(z_1) - E_0(z_2) = 0, \\
	& \mu(z_1) - \mu(z_2) = 0,
\end{array}
\right .
\label{1d15cw}
\end{equation}
we find the values $z_1$ and $z_2$ of the quantity $\bar z$, which determine the points of phase coexistence. 

Finally, the well-known relation $P V = k_B T \ln \Xi$, which connects the pressure, $P$, with the grand partition function, $\Xi$, allows us to derive the following equation of state for the model:
\begin{equation}
P v \beta_c = (\tau + 1) \left[\ln K_0(\bar z) - \frac{p_a}{2}\left(\frac{K_1(\bar z)}{K_0(\bar z)}\right)^2\right].
\label{1d16cw}
\end{equation} 
By utilizing the explicit relation \eqref{1d11cw} between the reduced density, $\bar n$, and the quantity $\bar z$ in \eqref{1d16cw}, we can express the equation of state in terms of pressure-temperature-density in a parametric form with $\bar{z}$ being the parameter. Following the same path we can express the chemical potential \eqref{1d10cw} in terms of the reduced density, $\bar n$.

In the following sections, we express the results by using dimensionless quantities:
\begin{align}\label{1d17cw}
	 \tilde P &= P v \beta_c \;\quad \text{ \--- the dimensionless pressure,} \nonumber \\
	 \tilde \mu &= \beta_c \mu \qquad \text{ \--- the dimensionless chemical potential.}  
\end{align}

\section{\label{sec:volume} Effect of changing the cell size}

In this Section, we study the influence of changing the reduced cell volume, $v^*$, on the phase behavior of the cell model. In the framework of the cell model the cell volume is a constant quantity. Moreover in the thermodynamic limit the average number of particles $\langle N \rangle \rightarrow \infty$, the total volume of the system $V \rightarrow \infty$, but $v = {\mathrm{const}}$. Thus our investigation means the quantitative comparison of the values that the reduced particle number density, $\bar n$, the chemical potential, $\tilde \mu$, and the pressure, $\tilde P$, can take, in the cases we set $v^*=1$, $v^*=2$, $v^*=5$, $v^*=10$, $v^*=15$. 

Taking a close look on the formula \eqref{1d11cw} one sees that changing the cell volume has no impact on the reduced density, $\bar n$. It happens because the special functions $K_m(\bar z)$ in the representation \eqref{1d6cw}
has no dependency on $v^*$. 

\begin{figure}[h!]
	\centering
	\includegraphics[width=0.49\textwidth]{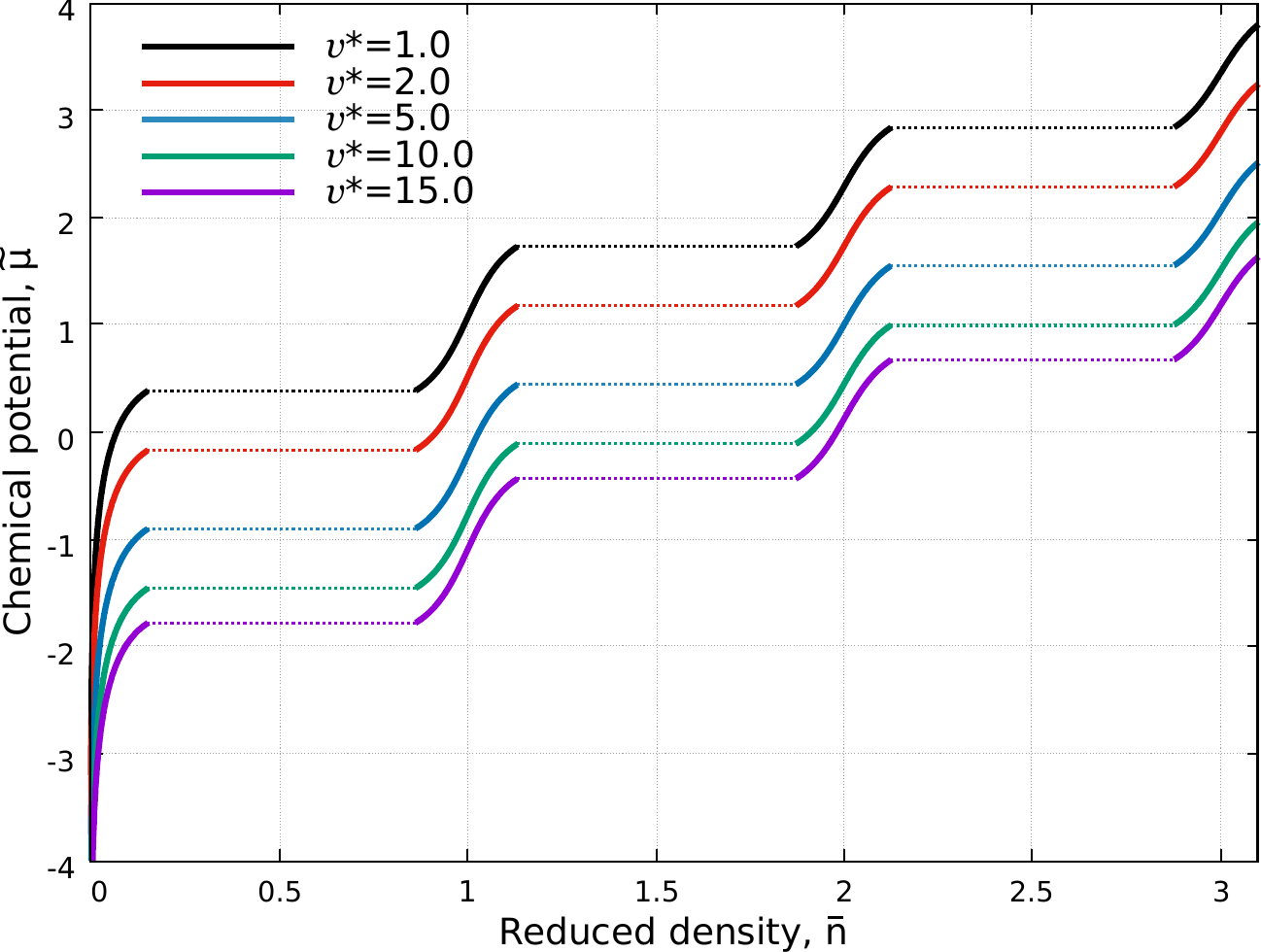} 
	\includegraphics[width=0.49\textwidth]{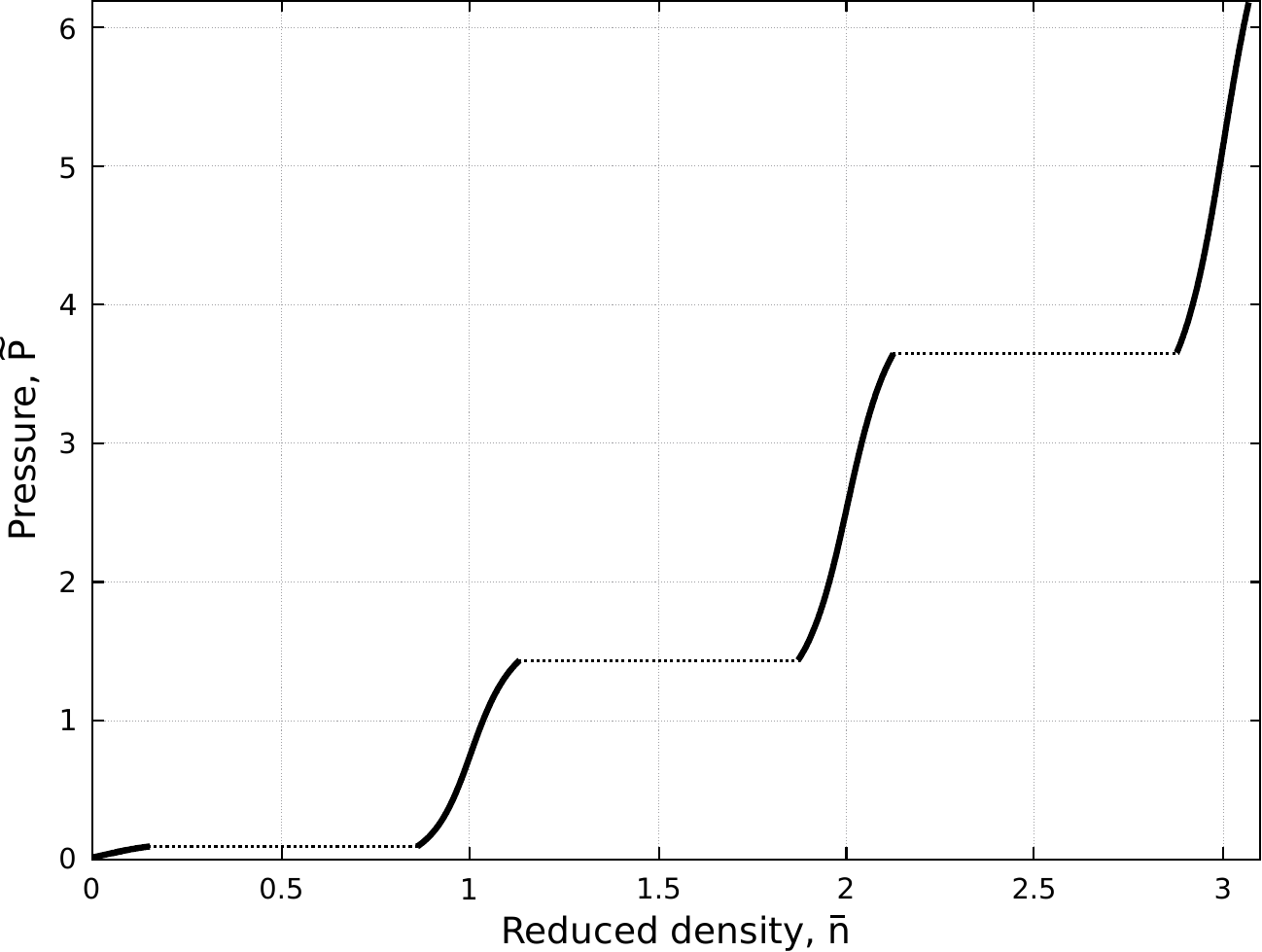}
	\vskip-3mm\caption{Part a: influence of varying the cell size, $v^*$, on isotherms of the chemical potential, $\tilde \mu$, as a function of the reduced density, $\bar n$: black line corresponds to $v^*=1$, red line \--- $v^*=2$, blue line \--- $v^*=5$, green line \--- $v^*=10$, purple line \--- $v^*=15$. Part b: isotherm of the pressure, $\tilde P$, as a function of the reduced density, $\bar n$, at arbitrary values of $v^*$. In both plots the reduced temperature is subcritical, $T/T_c = 0.8$, thus solid lines display regions of stable phases, thin dotted horizontal lines indicate density jumps during the first-order phase transitions. The repulsion to attraction ratio $f=1.2$. }\label{fig1}
\end{figure}

 Behavior of the chemical potential, $\tilde \mu$, versus the reduced density, $\bar n$, at various fixed values of the cell volume, $v^*$, is represented graphically in Figure~\ref{fig1}a.
The equation for the chemical potential \eqref{1d10cw} includes the cell volume explicitly. What we see regarding that, in Figure~\ref{fig1}a, is a parallel shift of the isotherms towards lower values of $\tilde \mu$ as the cell volume increases. In contrast to the chemical potential, the pressure, $\tilde P$, is insensitive to the cell volume change. Thus the isotherm of the pressure, $\tilde P$, represented in Figure~\ref{fig1}b remains the same for arbitrary values of $v^*$. Both plots in Figure~\ref{fig1} shows the case of the subcritical temperature, $T/T_c = 0.8$, and the repulsion to attraction ratio $f=1.2$.

\begin{table}[h!]
	\caption{Effect of varying the reduced cell volume, $v^*$, on critical values of the chemical potential, $\tilde \mu_c$, and the pressure, $\tilde P_c$, corresponding to the first three phase transitions in the sequence denoted by the superscripts (1), (2), (3) respectively. The repulsion to attraction ratio $f=1.2$.}
	\vskip-5mm 
	\tabcolsep4.5pt
	\label{tab1}

\begin{center}
\noindent{\footnotesize
	\begin{tabular}{|c|c|c|c|c|c|c|}
		\hline%
		\rule{0pt}{5mm}$v^*$   & $\tilde \mu^{(1)}_c$ & $\tilde \mu^{(2)}_c$ & $\tilde \mu^{(3)}_c$ & $\tilde P^{(1)}_c$ & $\tilde P^{(2)}_c$ & $\tilde P^{(3)}_c$   \\[1mm]%
		\hline%
		\rule{0pt}{4mm}1.0    & 0.379645		 & 1.886537		   & 3.106208 		 & 0.197746 	 &  1.719367 	 &  4.156744 	  \\%
		\hline%
		\rule{0pt}{4mm}2.0    & -0.313503		 & 1.173475		   & 2.383961 		 & 0.197746 	 &  1.719367 	 &  4.156744 	  \\%
		\hline%
		\rule{0pt}{4mm}5.0    & -1.229793		 & 0.230858		   & 1.429202 		 & 0.197746 	 &  1.719367 	 &  4.156744 	  \\%
		\hline%
		\rule{0pt}{4mm}10.0   & -1.922940		 & -0.482205	   & 0.706956 		 & 0.197746 	 &  1.719367 	 &  4.156744 	  \\%
		\hline%
		\rule{0pt}{4mm}15.0   & -2.328406		 & -0.899319	   & 0.284468 		 & 0.197746 	 &  1.719367 	 &  4.156744 	  \\%
		\hline
	\end{tabular} }
\end{center}
\end{table}
 We have already mentioned in Section~\ref{sec:model}, that the cell model under consideration exhibits multiple first-order phase transitions as the density and the chemical potential (or the pressure) are increased at a given temperature. Figure~\ref{fig1} covers the density region where we observe a series of three sequential phase transitions between four stable phases (thick solid lines). This phenomenon is literally displayed by three horizontal thin dotted lines — three sequential jumps of the order parameter, namely the reduced density.
As we fix the temperature each subsequent phase transition occurs at an increasingly higher density and chemical potential (Figure~\ref{fig1}a) or pressure (Figure~\ref{fig1}b).
Obviously, each phase transition displayed in Figure~\ref{fig1} terminates at a distinct critical point. Table~\ref{tab1} represents critical values of the chemical potential, $\tilde \mu_c$, and the pressure, $\tilde P_c$, in cases of different fixed values of the cell size, $v^*$.

\section{\label{sec:interaction} Impact of the interaction intensity}

In this section, we investigate how variations in the ratio of repulsive to attractive interaction intensities, defined as $f = g_r / g_a$, affect the phase behavior of the cell model. The parameter $f$ is one of only two adjustable parameters in the model required to obtain quantitative results (the other being the cell volume $v$, discussed previously in Section~\ref{sec:volume}). It is reasonable to expect that interaction intensities will significantly influence the model’s phase behavior. This is supported by the explicit presence of $g_r$ and $g_a$ in key expressions of the special functions~\eqref{1d6cw}, the chemical potential~\eqref{1d10cw}, the reduced density~\eqref{1d11cw}, and the pressure~\eqref{1d16cw}.

Quantitative and graphical results for the case  $f=1.2$ within the cell model with Curie-Weiss interaction based on the equation of state~\eqref{1d16cw} have already been presented in~\cite{kd122}. In the present work, we expand on that analysis by displaying pressure–temperature and temperature–density phase diagrams for values of $f$: $1.01$, $1.2$, $1.5$, $2.0$, and $5.0$ (see Figure~\ref{fig2}).

\begin{figure}[t!]
	
	\vspace{1.5cm}
	
	\centering
	\includegraphics[width=0.49\textwidth]{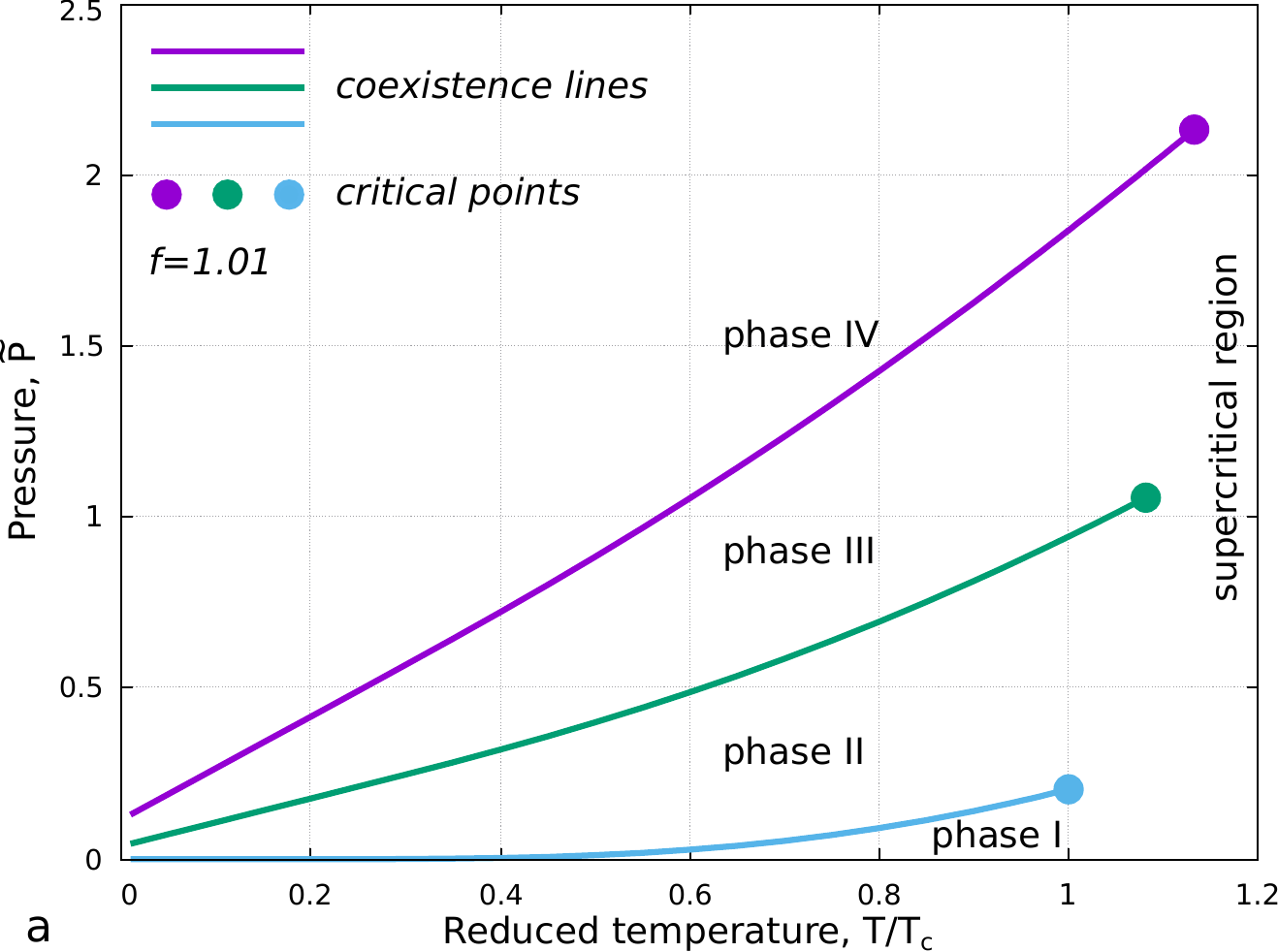} 
	\includegraphics[width=0.49\textwidth]{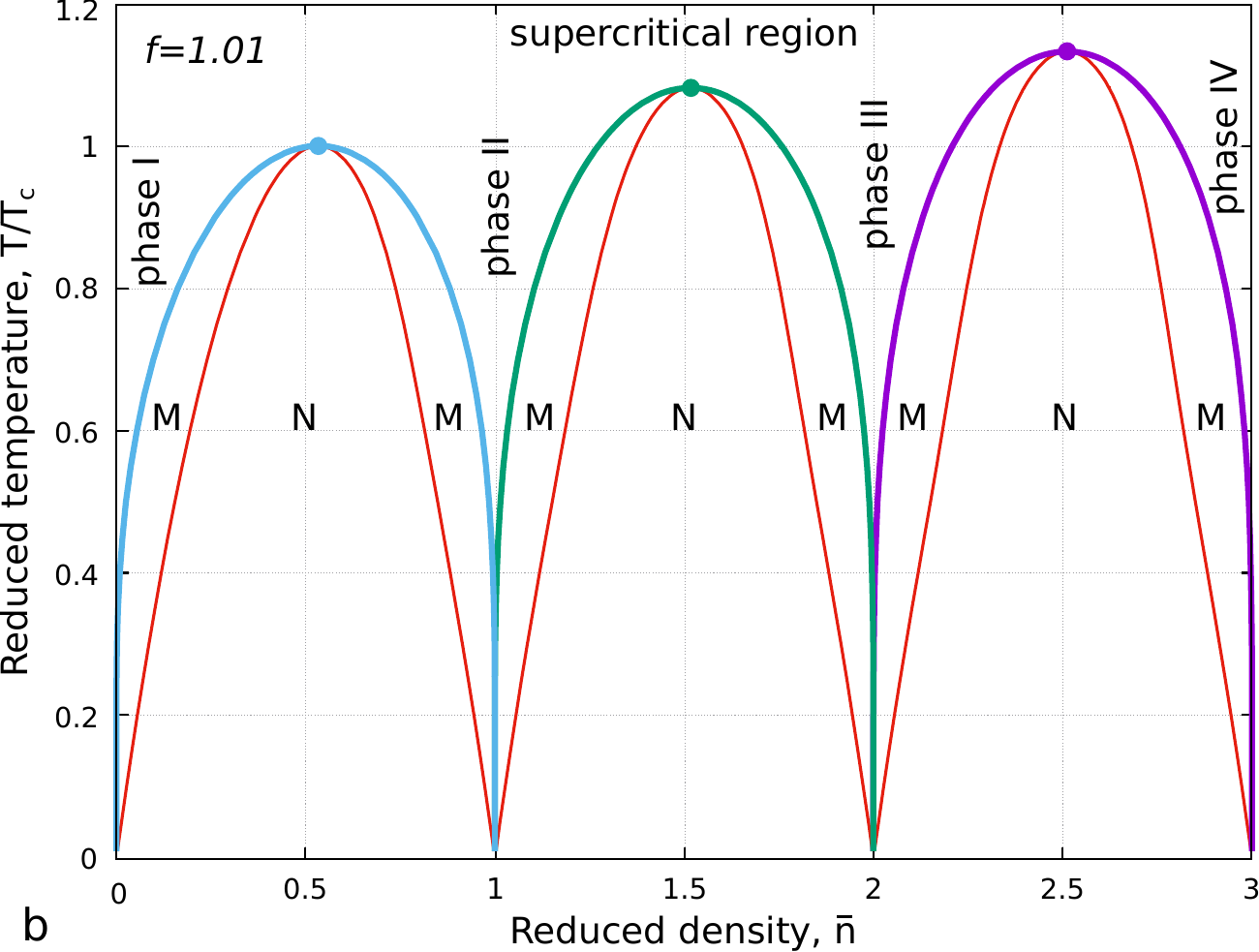}
	
	\vspace{1.5cm}
	
	\includegraphics[width=0.49\textwidth]{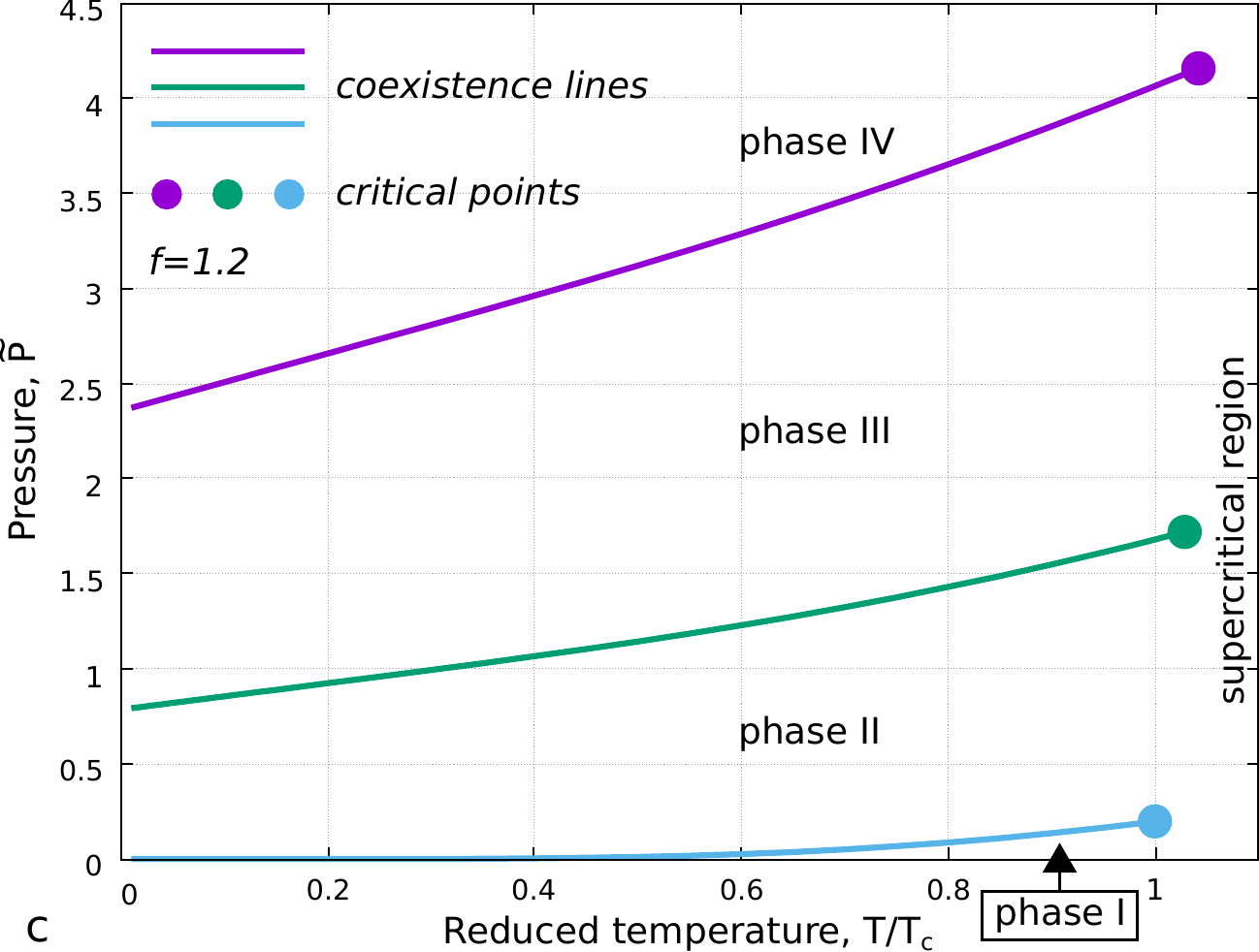} 
	\includegraphics[width=0.49\textwidth]{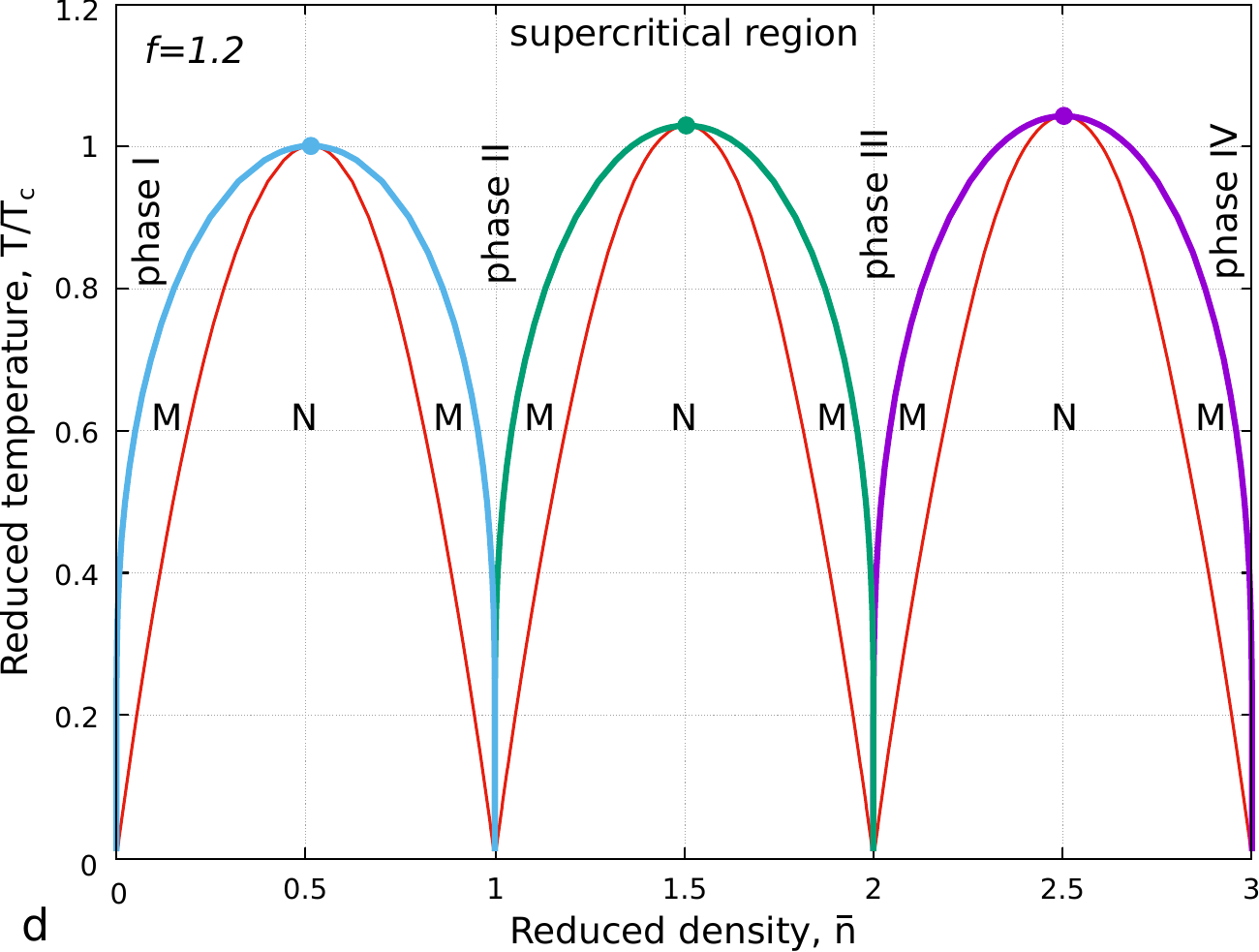}
	
	\vspace{1.5cm}
	
	\includegraphics[width=0.49\textwidth]{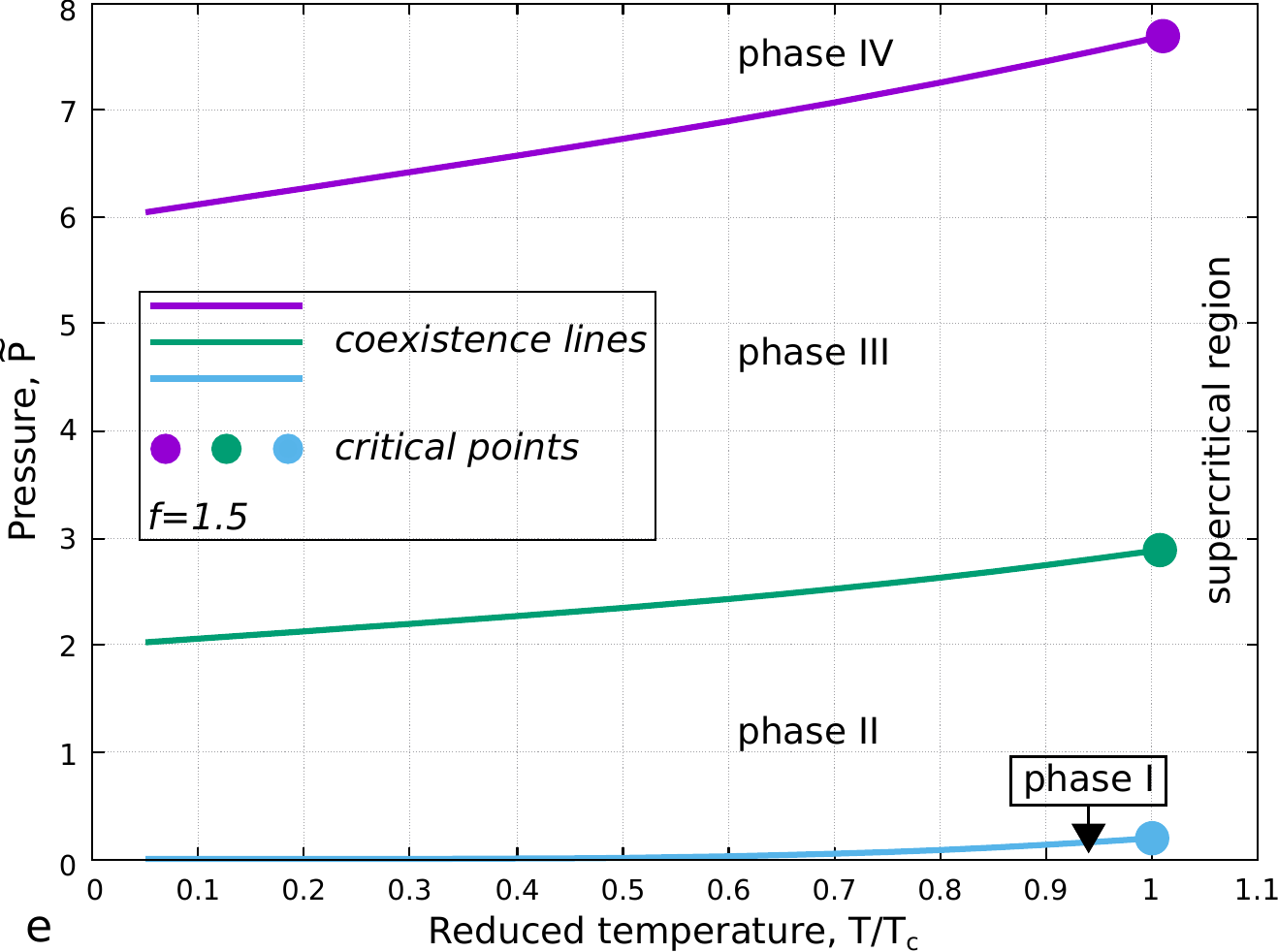} 
	\includegraphics[width=0.49\textwidth]{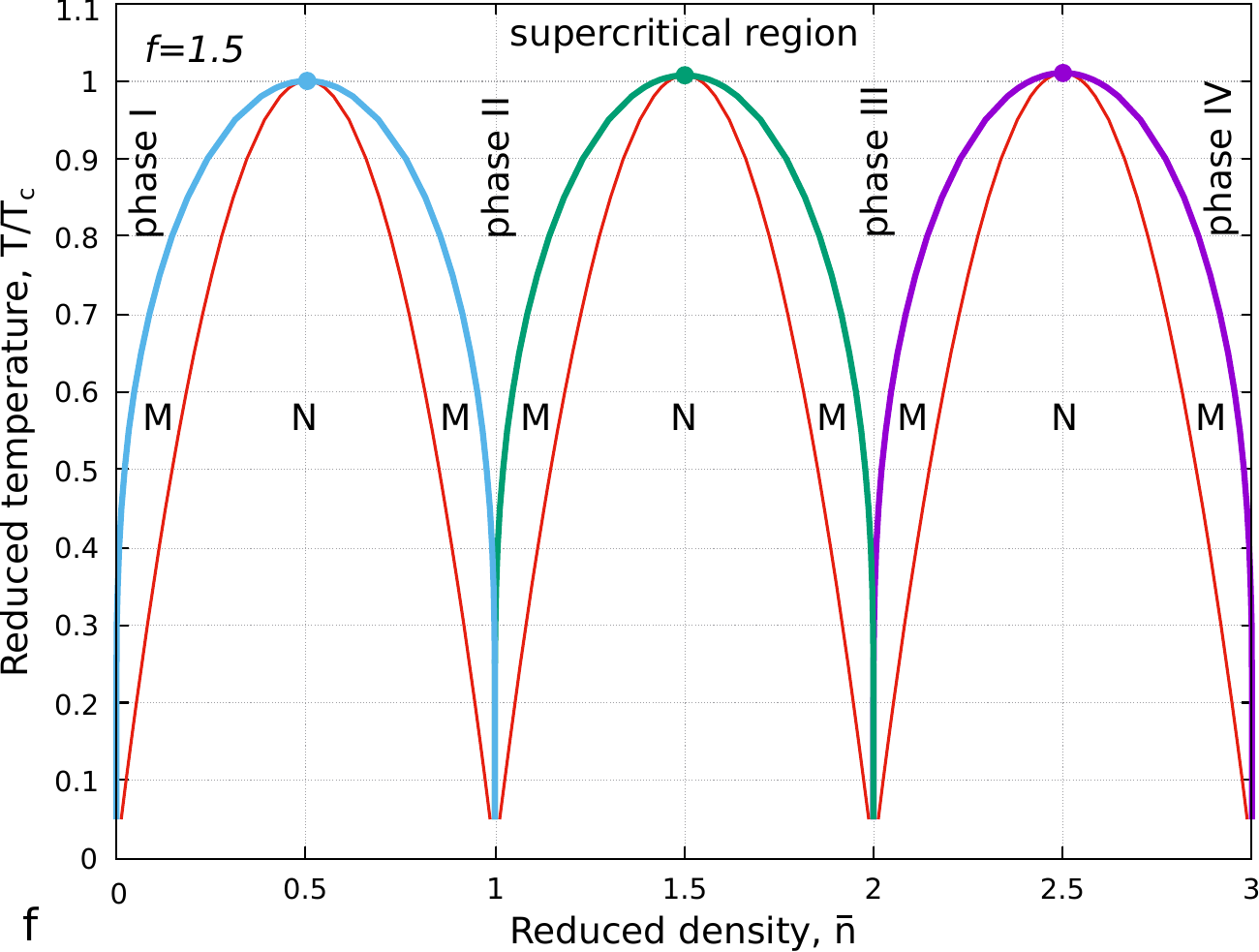}
	
	\vspace{1.5cm}
	
\end{figure}
\begin{figure}[t!]
	\centering
	\includegraphics[width=0.495\textwidth]{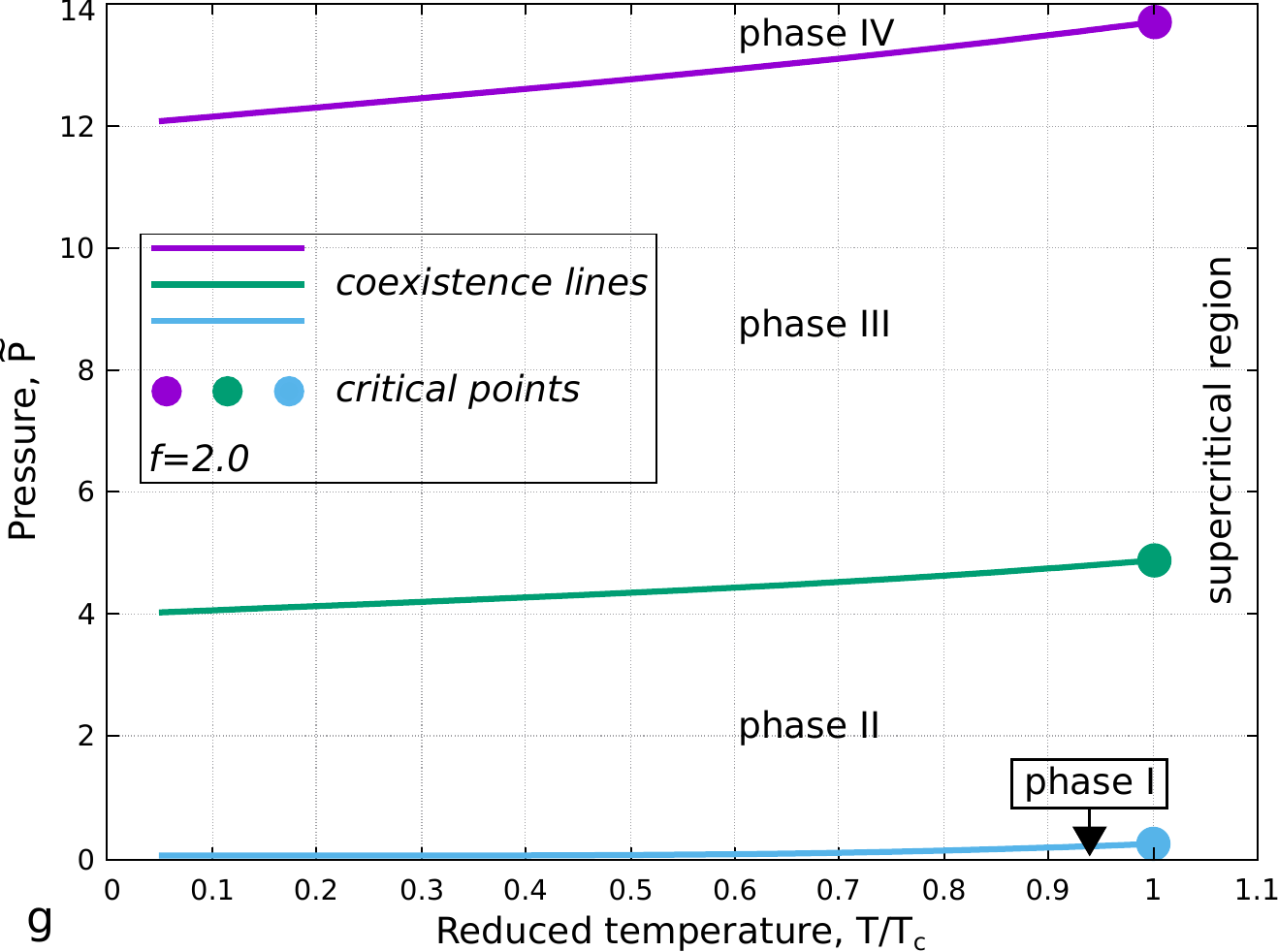} 
	\includegraphics[width=0.49\textwidth]{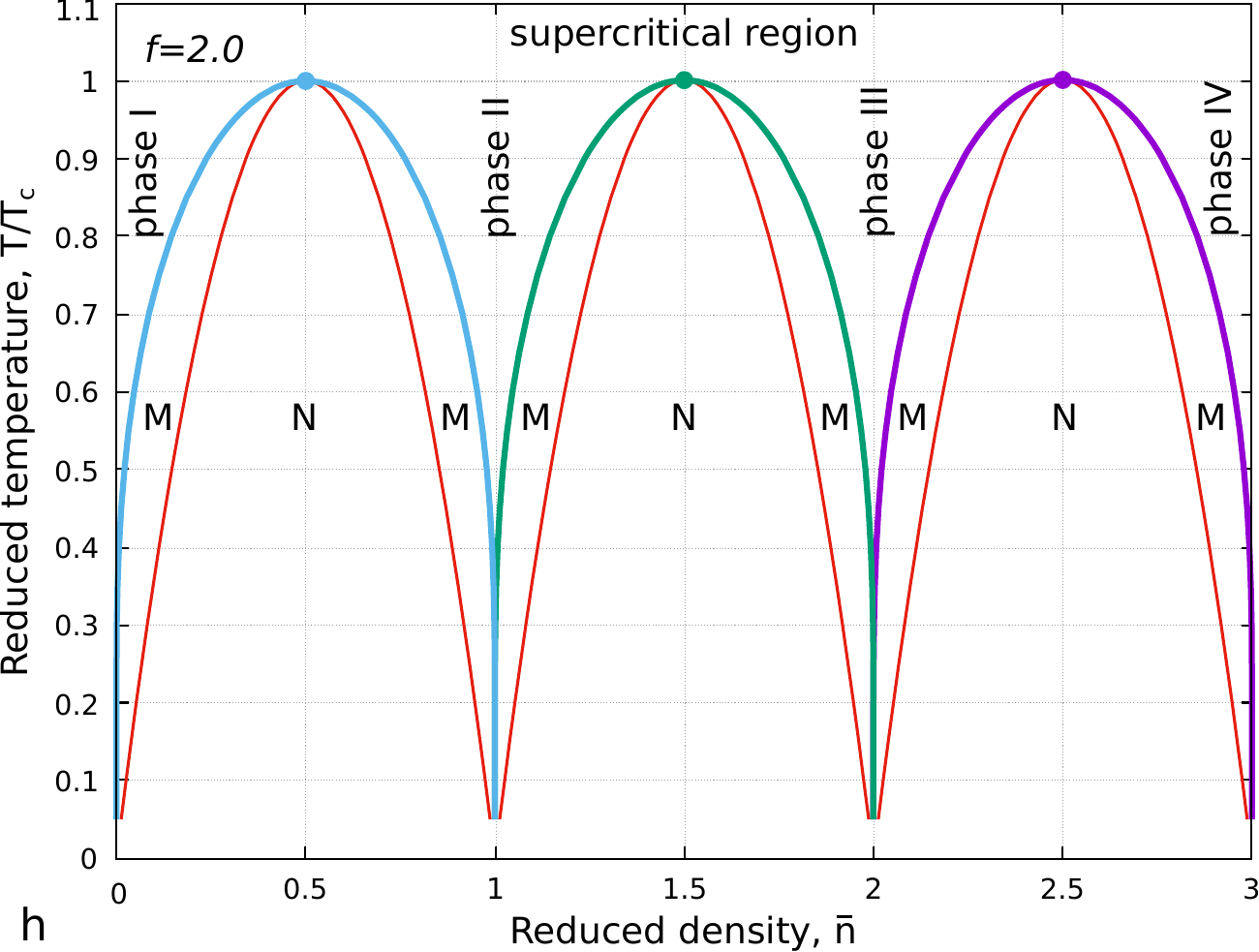}
	
	\vspace{1.5cm}
	
	\includegraphics[width=0.49\textwidth]{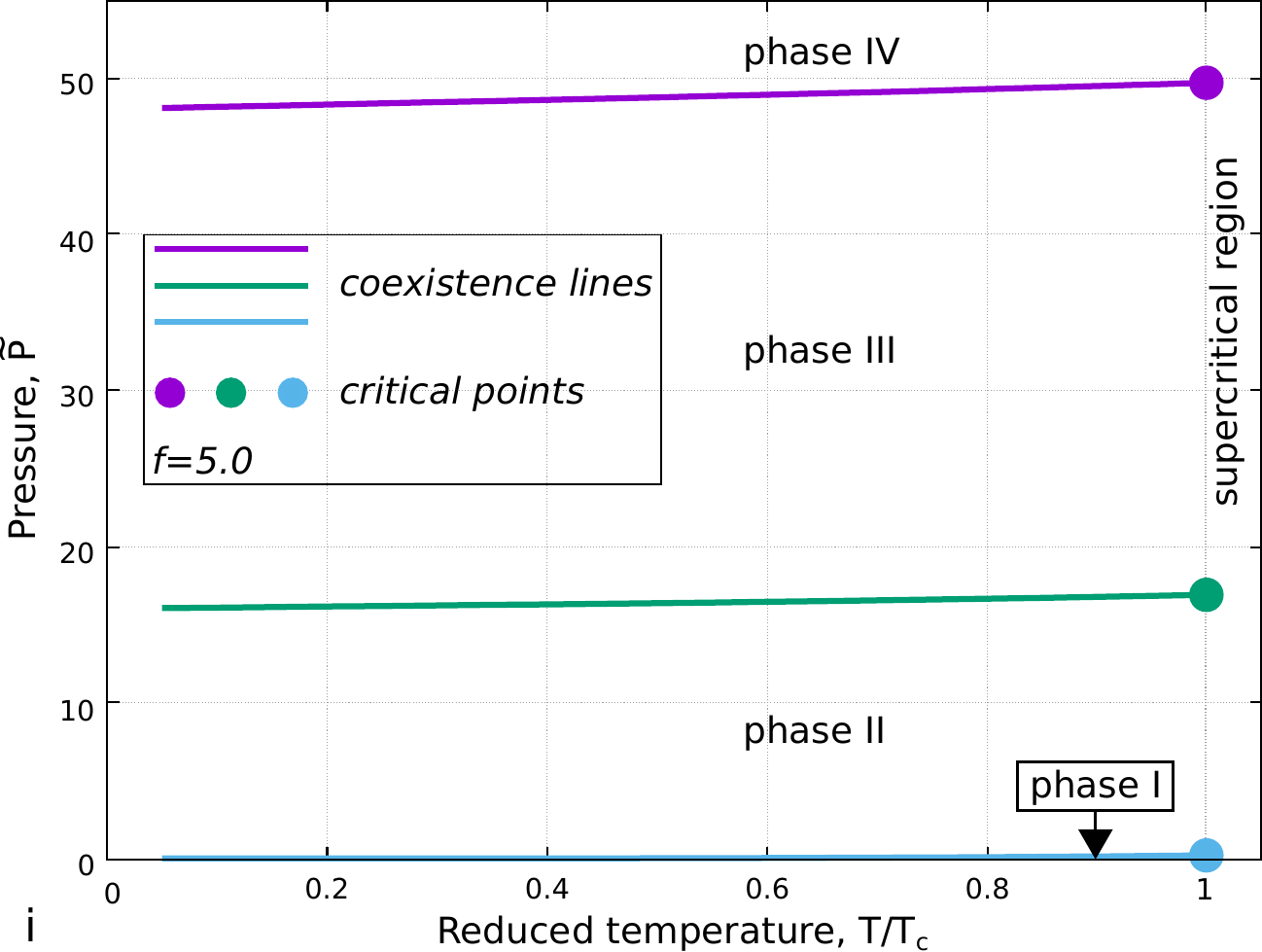} 
	\includegraphics[width=0.49\textwidth]{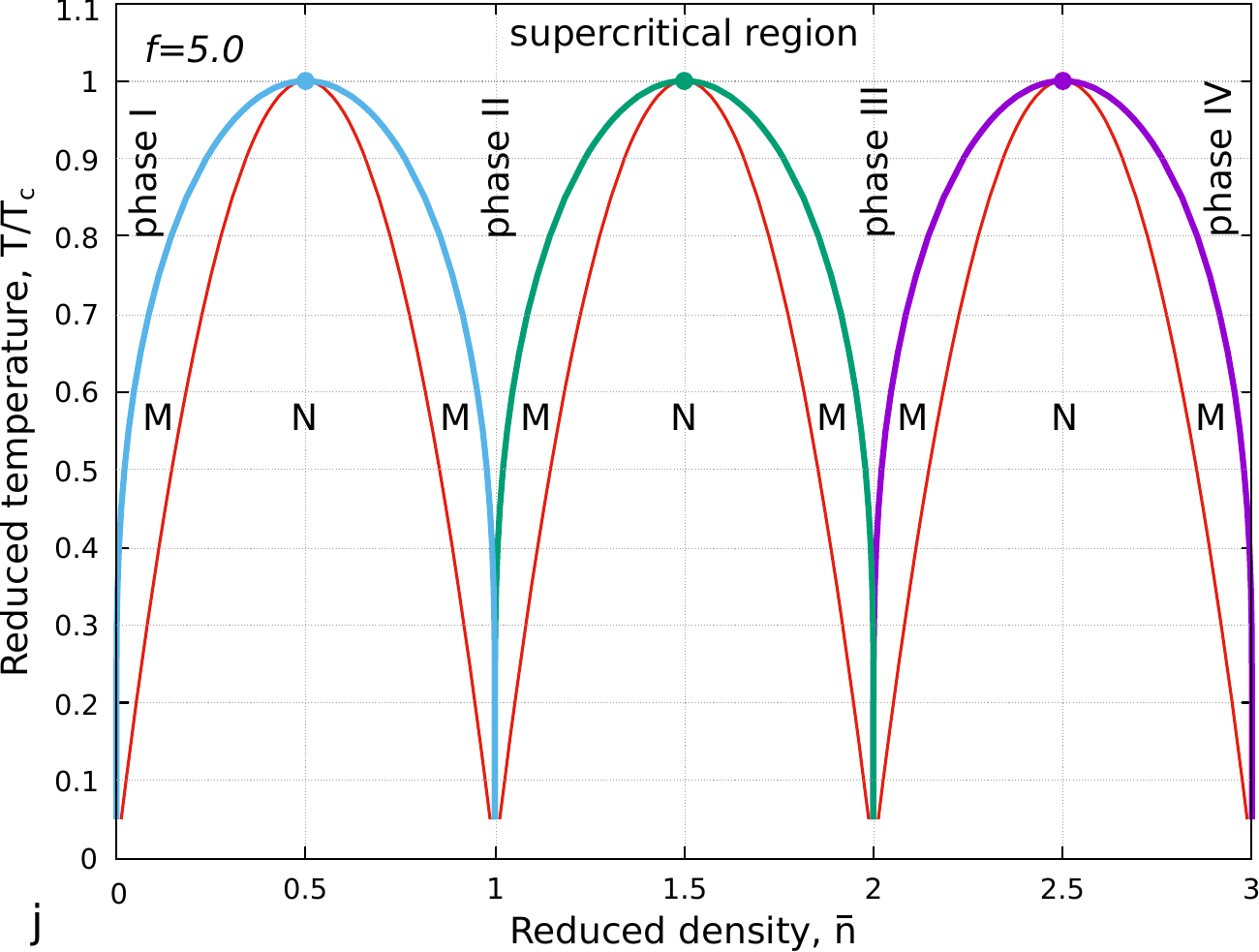}
	\caption{Phase diagrams in the pressure-temperature (plots a, c, e, g, i) and temperature-density (plots b, d, f, h, j) planes for the cell model with a Curie-Weiss-type interaction potential~\eqref{1d2cw}, in cases we set the repulsion to attraction ratio $f = 1.01$ (plots a, b), $f = 1.2$ (c, d), $f = 1.5$ (e, f), $f = 2.0$ (g, h), $f = 5.0$ (i, j).  All diagrams are valid for arbitrary cell sizes. 
		\\ Plots~(a, c, e, g, i): Phase diagram in the ($\tilde P,T/T_c$)-plane showing the first three first-order phase transitions and the corresponding phase coexistence lines among four distinct phases. Each coexistence line terminates at its critical point (solid circles) on the right-hand side.  
		\\ Plots~(b, d, f, h, j): Phase diagram in the ($T/T_c, \bar n$)-plane displaying spinodals (thin red lines) and phase coexistence lines: blue solid line indicates coexistence between Phase I and Phase II, green solid line — coexistence between Phase II and Phase III, and purple solid line — coexistence between Phase III and Phase IV. Regions marked by M and N denote metastable and unstable states, respectively.	
	}\label{fig2}
\end{figure}

In Figure~\ref{fig2}, plots a, c, e, g, and i show the phase equilibria in the pressure–temperature projection. These diagrams span temperatures from as low as $0.02 T_c$ (Figure~\ref{fig2}a and Figure~\ref{fig2}c) and $0.05 T_c$ (Figure~\ref{fig2}e, Figure~\ref{fig2}g, and Figure~\ref{fig2}i) up to the supercritical region. Each plot features three phase coexistence lines, corresponding to a sequence of first-order phase transitions: blue line — transition between Phase~I and Phase~II, green line — transition between Phase~II and Phase~III, purple line — transition between Phase~III and Phase~IV. Each coexistence line terminates at its corresponding critical point.

Across all values of the parameter $f$, the first phase transition (blue line) consistently occurs at relatively low reduced pressures, $\tilde P$, ranging from $0.0$ to $P^{*(1)}_c \approx 0.2$ (see Table~\ref{tab_2cw}). The second phase transition (green line) always occurs at higher pressures than the first, and the pressure range at which Phase~II~–~Phase~III coexistence appears increases with the repulsion-to-attraction ratio. For example, this $\tilde P$ range is $2.02...2.89$ for $f=1.5$ while for $f=5.0$ it shifts to $16.03..16.89$. The third phase transition (purple line) follows a similar trend but occurs at much higher pressures. Note that none of the phase diagrams in Figure~\ref{fig2} display a triple point.

Table~\ref{tab_2cw} presents the critical point parameters for the first three phase transitions in the sequence, evaluated for various values of the repulsion-to-attraction ratio, $f$. At low $f$ values ($1<f<3$), each subsequent phase transition is associated with a higher critical temperature. However, this trend gradually disappears as $f$ increases. For $f>3$, all critical points from the sequence occur at the same critical temperature (with precision of many decimal places). This behavior is also clearly visible in the phase diagrams shown in Figure~\ref{fig2}.

In Figure~\ref{fig2}, plots b, d, f, h, and j display the phase equilibria in the temperature-density projection. As in corresponding pressure-temperature diagrams, each plot features three binodals, representing a sequence of first-order phase transitions: blue line — the transition between Phase~I and Phase~II, green line — the transition between Phase~II and Phase~III, purple line — the transition between Phase~III and Phase~IV. In addition, these diagrams include regions of metastable and unstable states, bounded by spinodal curves (thin red lines). 
\begin{table}[h!]
	\caption{Critical values of the quantity $p_a$ and the variable $\bar{z} $, as well as the parameters of the critical points — the critical temperature, $T_c$, the critical reduced density, $\bar n_c$, the critical pressure, $\tilde P_c$, and the critical chemical potential, $\tilde \mu_c$ — corresponding to first three phase transitions in the sequence (denoted by the superscript ($n$)) for different repulsion-to-attraction ratio $f$ ($v^* = 5.0$).}
	\vskip-5mm
	\tabcolsep4.5pt
	\label{tab_2cw}
	
	\begin{center}
		\noindent{\footnotesize\begin{tabular}{|c|c|c|c||c|c|c|c|}
				\hline%
				{\rule{0pt}{5mm}$f$} & $(n)$ & $p_{ac}^{(n)}$ & $ \bar{z}_c^{(n)}$  &  $T_c^{(n)}$ & $\bar n_c^{(n)}$ &
				$\tilde P_c^{(n)}$ & $\tilde \mu_c^{(n)}$ \\[1mm]%
				\hline
				\rule{0pt}{4mm}	& 1 & 3.8254 & 2.0175 & $T_c$ & 0.5355 & 0.2048 & $-1.6405$ \\%
				1.00001 & 2 & 3.5199 & 6.0275 & 1.0868$T_c$ & 1.5191 & 1.0262 & $-1.0097$ \\%
				& 3 & 3.3471 & 9.5022 & 1.1429$T_c$ & 2.5128 & 2.0407 & $-0.5920$ \\%
				\hline
				& 1 & 3.8333 & 2.0355 & $T_c$ & 0.5338 & 0.2042 & $-1.6200$ \\%
				1.01 & 2 & 3.5443 & 6.1135 & 1.0815$T_c$ & 1.5177 & 1.0562 & $-0.9467$ \\%
				& 3 & 3.3842 & 9.6763  & 1.1327$T_c$ & 2.5116 & 2.1314 & $-0.4905$ \\%
				\hline
				& 1 & 3.8612 & 2.1093 & $T_c$ & 0.5277 & 0.2023 & $-1.5379$ \\%
				1.05 & 2 & 3.6286 & 6.4468 & 1.0641$T_c$ & 1.5134 & 1.1834 & $-0.6960$ \\%
				& 3 & 3.5084 & 10.3316 & 1.1005$T_c$ & 2.5082 & 2.5191 & $-0.0855$ \\%
				\hline
				& 1 & 3.8890 & 2.2039 & $T_c$ & 0.5219 & 0.2004 & $-1.4351$ \\%
				1.1 & 2 & 3.7097 & 6.8426 & 1.0483$T_c$ & 1.5097 & 1.3541 & $-0.3854$\\%
				& 3 & 3.6230 & 11.0781 & 1.0734$T_c$ & 2.5056 & 3.0420 & 0.4195 \\%
				\hline
				& 1 & 3.9282 & 2.3983 & $T_c$ & 0.5139 & 0.1977 & $-1.2298$ \\%
				1.2 & 2 & 3.8185 & 7.5829 & 1.0287$T_c$ & 1.5056 & 1.7194 & 0.2309 \\%
				& 3 & 3.7700 & 12.4174 & 1.0420$T_c$ & 2.5030 & 4.1567 & 1.4292 \\%
				\hline
				& 1 & 3.9531 & 2.5963 & $T_c$ & 0.5090 & 0.1961 & $-1.0252$\\%
				1.3 & 2 & 3.8841 & 8.2772 &  1.0178$T_c$ & 1.5034 & 2.1018 & 0.8432\\%
				& 3 & 3.8552 & 13.6332 & 1.0254$T_c$ & 2.5018 & 5.3174 & 2.4394 \\%
				\hline
				& 1 & 3.9796 & 2.9962 & $T_c$ & 0.5039 & 0.1944 & $-0.6184$ \\%
				1.5 & 2 & 3.9509 & 9.5867 & 1.0073$T_c$ & 1.5014 & 2.8883 & 2.0604 \\%
				& 3 & 3.9395 & 15.8739 & 1.0102$T_c$ & 2.5007 & 7.6917 & 4.4578 \\%
				\hline
				& 1 & 3.9910 & 3.3974  & $T_c$ & 0.5017 & 0.1937 & $-0.2143$ \\%
				1.7 & 2 & 3.9786 & 10.8402 & 1.0031$T_c$ & 1.5006 & 3.6852 & 3.2708 \\%
				& 3 & 3.9738 & 17.9881 & 1.0043$T_c$ & 2.5003 & 10.0889 & 6.4708 \\%
				\hline
				& 1 & 3.9973 & 3.9988 & $T_c$ & 0.5005 & 0.1933 & 0.3887 \\%
				2.0 & 2 & 3.9937 & 12.6747 & 1.0009$T_c$ & 1.5002 & 4.8851 & 5.0787 \\%
				& 3 & 3.9923 & 21.0603 & 1.0013$T_c$ & 2.5001 & 13.6923 & 9.4817 \\%
				\hline
				& 1 & 4.0000 & 6.0000 & $T_c$ & 0.5000 & 0.1932 & 2.3905\\%
				3.0 & 2 & 3.9999 & 18.6926 & 1.00001$T_c$ & 1.5000 & 8.8862 & 11.0835 \\%
				& 3 & 3.9999 & 31.0976 & 1.00002$T_c$ & 2.5000 & 25.6970 & 19.4889\\%
				\hline
				& 1 & 4.0000 & 10.0000 & $T_c$ & 0.5000 & 0.1931 & 6.3906 \\%
				5.0 & 2 & 4.0000 & 30.6931  & $T_c$ & 1.5000 & 16.8863 & 23.0837 \\%
				& 3 & 4.0000 & 51.0986 & $T_c$ & 2.5000 & 49.6972 & 39.4892 \\%
				\hline
				& 1 & 4.0000 & 20.0000 & $T_c$ & 0.5000 & 0.1931 & 16.3906 \\%
				10.0 & 2 & 4.0000 & 60.6931 & $T_c$ & 1.5000 & 36.8863 & 53.0837 \\%
				& 3 & 4.0000 & 101.0986 & $T_c$ & 2.5000 & 109.6972 & 89.4892 \\[0mm]%
				\hline
		\end{tabular}}
	\end{center}
\end{table}

A closer examination of the temperature–density phase diagrams reveals a consistency in the location of the coexistence curves with respect to the reduced particle number density $\bar n$, regardless of the values of $f$. Specifically, the critical points occur at approximately the same density values: at the first critical point $\bar n \approx 0.5$, at the second $\bar n \approx 1.5$, and at the third $\bar n \approx 2.5$ (see also Table~\ref{tab_2cw}). Depending on the temperature, the stable phases are observed in the following density ranges: Phase~I is in the reduced density range between $0$ and $0.5$, Phase~II between $0.5$ and $1.5$, Phase~III between $1.5$ and $2.5$, and Phase~IV between $2.5$ and $3.5$. In the context of the cell model, the reduced density represents the average number of particles per cell.

\section{Conclusions}

In the present study, we establish the roles of the two key parameters used in formulating the cell model.
The first parameter is the volume of the cubic cell, $v$, into which the total system volume is conventionally divided. We have shown that this parameter has no influence on the phase transition characteristics. In particular, the equation of state in pressure-temperature-density terms retains a universal form, and the coordinates of the critical point remain unchanged. However, this parameter has impact on the values of the chemical potential, $\bar \mu$. 

The second parameter significantly influences the equation of state, the critical point parameters, and several other important characteristics. This parameter represents the ratio of repulsive to attractive intensities between particles, $f$, which, in our case, is a Curie-Weiss-type interaction~\eqref{1d2cw}. According to~\cite{r170}, to ensure the stability of interaction, this ratio cannot be less than one, $f \geq 1$. That is, attraction cannot exceed repulsion, otherwise, the system will tend to collapse. In the present study we analyzed the range $1<f\leq 10$. Our results confirm that increasing the repulsion-to-attraction ratio lowers the critical temperature. However, when this ratio exceeds 3, $f > 3$, the critical temperature saturates and no longer decreases. In this case, the critical temperatures of each phase transition in the sequence are equal. As $f$ increases, the pressures at which the phase transitions occur also rise, particularly for the second and third transitions. However, as $f$ increases, the critical points continue to occur at approximately the same reduced densities as they do at lower values of $f$. This consistency reflects the nature of the reduced density as the average number of particles per cell in the model, which remains almost unaffected by changes in~$f$.

In summary, varying the cell volume, $v$, does not affect the pressure-temperature-density phase diagrams. In contrast, increasing the ratio of repulsive to attractive interaction intensities between particles, $f$, influences the phase diagrams quantitatively, for example, by expanding the pressure range over which stable phases exist or by decreasing the critical temperature until it saturates, resulting in equal critical temperatures for all phase transitions in the sequence. However, changes in $f$ do not lead to qualitative alterations in the phase behavior of the cell model: phenomena such as the disappearance of phase transitions or the emergence of triple points are not observed.

\section*{Acknowledgement}

Yuriy O. Plevachuk is supported by the Office of Government of Slovakia, project no. 09I03-03-V01-00047.

\end{document}